\documentclass[letterpaper, 10 pt,  conference]{ieeeconf}  

\usepackage{graphicx} 
\usepackage{amsmath,bm,times} 
\usepackage{amssymb}  
\usepackage{mathabx}
\usepackage{subfigure}
\usepackage{cite}
\usepackage{soul}
\usepackage{balance}
\usepackage{tabularx}
\usepackage{url}

\usepackage[linesnumbered, ruled, vlined]{algorithm2e}
\usepackage{algorithmicx}

\newtheorem{assumption}{Assumption}

\newtheorem{proposition}{Proposition}
\newtheorem{remark}{Remark}
\newtheorem{theorem}{Theorem}
\newtheorem{definition}{Definition}

\newtheorem{example}{Example}

\DeclareMathOperator*{\argmax}{arg\,max}

\title{\LARGE \bf Robust Decision-Making in Finite-Population Games}%
\author{Shinkyu Park and Lucas C. D. Bezerra
  \thanks{This work was supported by funding from King Abdullah University of Science and Technology (KAUST).}
  \thanks{The authors are with Electrical and Computer Engineering, King Abdullah University of Science and Technology (KAUST), Thuwal 23955, Saudi Arabia. {\tt \{shinkyu.park, lucas.camaradantasbezerra\}@kaust.edu.sa}}
}

\IEEEoverridecommandlockouts
\begin{document}

\maketitle

\begin{abstract}
  We study the robustness of an agent decision-making model in finite-population games, with a particular focus on the \textit{Kullback-Leibler Divergence Regularized Learning (KLD-RL)} model. Specifically, we examine how the model’s parameters influence the impact of various sources of noise and modeling inaccuracies---factors commonly encountered in engineering applications of population games---on agents' decision-making. Our analysis provides insights into how these parameters can be effectively tuned to mitigate such effects. Theoretical results are supported by numerical examples and simulation studies that validate the analysis and illustrate practical strategies for parameter selection.
\end{abstract}

\section{Introduction} \label{sec:intro}
The population game and evolutionary dynamics framework provides a powerful foundation for modeling and analyzing repeated strategic interactions among a population of decision-making agents \cite{Sandholm2010-SANPGA-2}. In this work, we consider a class of population games in which agents' payoffs---driving their strategy selection---are determined by a dynamic payoff mechanism. Specifically, payoff vectors evolve according to a dynamical system model that depends on both the internal state of the game and the population's state, which reflects agents' strategy selections. Our primary focus is on a particular subclass of these games, known as \textit{task allocation games} \cite{10383344, 9561809}, where agents interact strategically to accomplish a given set of tasks.

The conventional framework relies on several key assumptions: $1)$ payoff observations are noise-free, $2)$ switching strategies incurs no delay, and $3)$ the population size is infinite. However, these assumptions are often unrealistic in engineering applications of population games.
We specifically examine decision-making processes in finite-population games, where agents select strategies based on noisy payoff estimates influenced by their own past actions, and where strategy updates are subject to time delays. A central challenge posed by noise and modeling errors is the increased variability in agents' decision-making, which can hinder convergence to the optimal strategy selection.

Our study focuses on identifying a decision-making model from the population game literature that is robust to noise and modeling errors, and on understanding how its parameters can be optimized to reduce the impact of such uncertainties while preserving equilibrium learning. To address this, we investigate the potential of the \textit{Kullback-Leibler Divergence Regularized Learning (KLD-RL)} model \cite{10966158} as a solution. Through theoretical analysis, we examine how careful tuning of the model's parameters can mitigate the effects of noise and enhance stability of agent decision-making behavior. We validate these findings with simulation studies, demonstrating how parameters can be optimized.

This work builds upon and extends recent studies in multi-agent games \cite{10156113, 10886571, 9993385, 5137415, BRAVO201741, Obando2016}, which analyze the impact of noise in payoff observations \cite{10886571, 9993385, 5137415, BRAVO201741} and time delays \cite{10156113, Obando2016}. Unlike our earlier works \cite{10156113, 10886571}, which focus primarily on the Smith model, here we examine the robustness of the KLD-RL model. As discussed in Section~\ref{sec:formulation}, this model exhibits a strong notion of \textit{passivity}, allowing agents to maintain robustness in the presence of perturbations. Our results therefore suggest that KLD-RL is particularly well-suited for realistic environments, where noise, modeling errors, and finite agent populations play a critical role.


Meanwhile, the conventional population game literature often investigates the perturbed best response model, in which payoffs are subject to random noise \cite{52177154-daeb-3f75-b380-9f9a398bed3c, HOFBAUER200747}. In contrast to our study, these works assume that the noise affecting the payoff vector is independent of both the payoffs themselves and the population state, which reflects the agents’ strategy choices. Our framework, by comparison, explicitly accounts for noise and modeling inaccuracies that depend on these variables.

The paper is organized as follows: Section~\ref{sec:formulation} introduces preliminaries on evolutionary dynamics and the class of population games considered in this work, followed by the problem formulation. Section~\ref{sec:simulation} presents an illustrative example with supporting simulations. In Section~\ref{sec:analysis}, we present analytical results that explain the robustness of the KLD-RL model and demonstrate how its parameters can be tuned to mitigate the effects of noise and delays in agents' decision-making. Finally, Section~\ref{sec:conclusions} concludes the paper with a summary and outlines directions for future research.

\textit{Notation:} Given a positive integer $n$, let $\mathbb R^n$ denote the spaces of $n$-dimensional real vectors, $\mathbb R_{\geq 0}^n$ the space of $n$-dimensional vectors with nonnegative entries, and $\mathbb R_{\geq 0}$ the set of nonnegative real numbers. For notational convenience, given positive integers $n$ and $N$, we define the finite set $\mathbb X^N = \{ (X_1^N, \cdots, X_n^N) \in \{0, 1/N, \cdots, 1 \}^n \,|\, \sum_{i=1}^n X_i^N = 1 \}$. Its infinite counterpart, obtained as $N \to \infty$, is the continuous simplex $\mathbb X = \{ (x_1, \cdots, x_n) \in [0,1]^n \,|\, \sum_{i=1}^n x_i = 1 \}$. Note that $\mathbb X^N$ is a subset of $\mathbb X$ for every positive integer $N$. Throughout the paper, we use the Euclidean norm $\| \cdot \|_2$ and the maximum norm $\| \cdot \|_\infty$. Additionally, given a vector $a$ and a vector-valued function $\mathcal A$, the notation $a_i$ and $\mathcal A_i$ refers to the $i$-th component of $a$ and $\mathcal A$, respectively.

\section{Preliminaries and Problem Description} \label{sec:formulation}

\begin{figure}
  \center
  \includegraphics[trim={0.0in 0.0in 0.0in 0.0in}, clip ,width=3.1in]{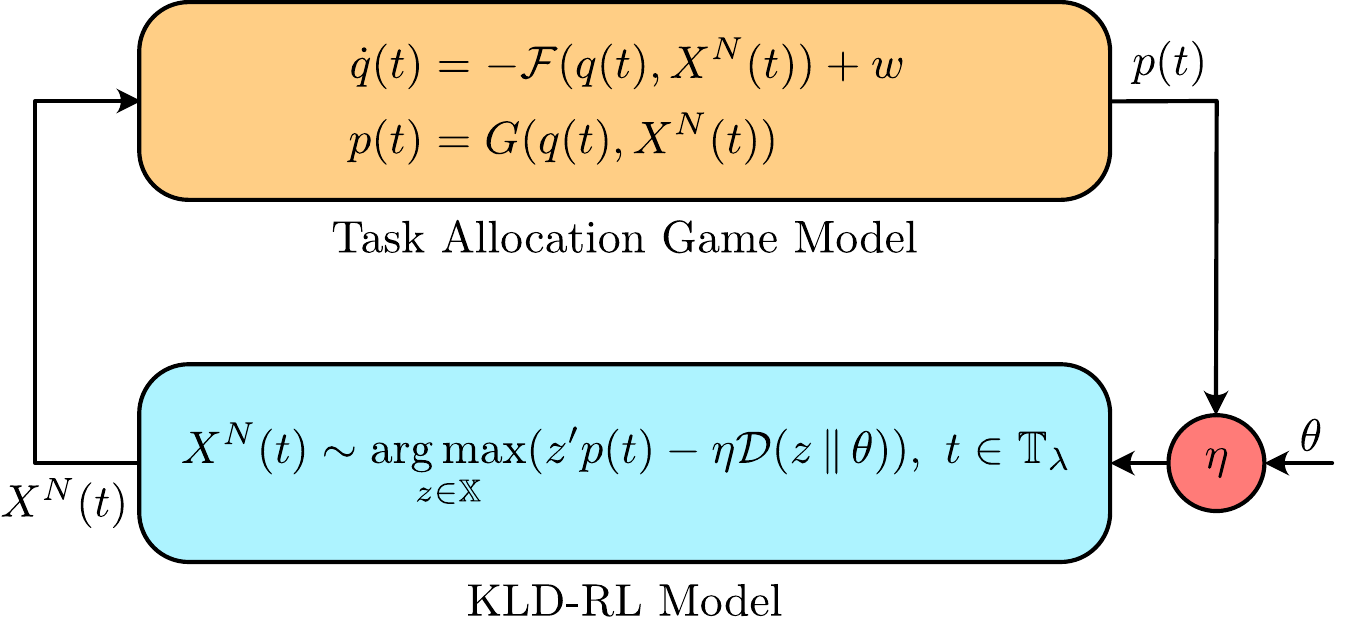}
  \caption{A feedback diagram consisting of the task allocation game model \eqref{eq:task_allocation_games} and the KLD-RL model \eqref{eq:kld_rl}. Each agent engaged in the game receives opportunities to revise its strategy at the arrival times $\mathbb T_\lambda$ of its associated Poisson process with rate parameter $\lambda$. The parameter $\eta$ in the KLD-RL model regulates the trade-off in decision-making between the payoff vector $p(t)$ and the reference distribution $\theta$.}
  \label{fig:feedback_diagram}
  \vspace{-1.0em}
\end{figure}

\subsection{Evolutionary Dynamics and Task Allocation Games}
Consider a finite population of $N$ agents engaged in a multi-agent game \cite{Sandholm2010-SANPGA-2}. Let the population state at time $t \geq 0$ be represented by the vector $X^N(t) = (X^N_1(t), \cdots, X^N_n(t)) \in \mathbb X^N$, where each component $X_i^N(t)$ denotes the fraction of agents selecting the $i$-th strategy among $n$ available options. The agents' strategy choices determine the payoff vector $p(t) = (p_1(t), \cdots, p_n(t)) \in \mathbb R^n$, according to an underlying game model, where each $p_i(t)$ represents the payoff received by agents selecting strategy~$i$. Due to the finite population size, each time an agent switches its strategy, the population state $X^N(t)$ undergoes a discrete \textit{jump} to a new state in the finite set $\mathbb X^N$. As illustrated in Fig.~\ref{fig:feedback_diagram}, the payoff vector and population state are jointly determined by a closed-loop system composed of the game dynamics and the decision-making model.

In population games, each agent has the opportunity to revise its strategy at the arrival times of an independent and identically distributed (i.i.d.) Poisson process with rate parameter $\lambda$. When such a revision opportunity occurs, the agent uses a \textit{strategy revision protocol} $\rho_{ji}: \mathbb R^n \times \mathbb X \to \mathbb R_{\geq 0}$ to determine the probability of switching from its current strategy~$j$ to a new strategy~$i$, based on the current payoff vector $p(t)$ and population state $X^N(t)$.

A well-known example of such a protocol is the \textit{Smith protocol}, defined as follows: for all $i,j \in \{1, \cdots, n\}$,
\begin{align} \label{eq:smith_protocol}
  \rho_{ji}^{\text{\tiny Smith}}(p, X^N)
  = \begin{cases}
    \varrho [p_i - p_j]_+ & \text{if } i \neq j \\
    1 - \sum_{\substack{i=1\\ i\neq j}}^n \varrho [p_i - p_j]_+ & \text{otherwise}, \\
  \end{cases}
\end{align}
where $\varrho$ is a positive constant that ensures \eqref{eq:smith_protocol} defines a valid probability distribution. Another revision protocol, which we focus on in this work, is the \textit{Kullback-Leibler Divergence Regularized Learning (KLD-RL)} protocol \cite{10966158}, defined as follows: for all $i,j \in \{1, \cdots, n\}$,
\begin{align} \label{eq:kld_rl_i}
  \rho_{ji}^{\text{\tiny KLD-RL}}(p, X^N) = C_{i}^{\eta,\theta} (p) = \frac{\theta_i \exp ( \eta^{-1} p_i )}{\sum_{l=1}^{n} \theta_l \exp ( \eta^{-1} p_l )},
\end{align}
where $\eta > 0$ and $\theta = (\theta_1, \cdots, \theta_n) \in \mathbb X$ is a \textit{reference distribution} over strategies.
Equivalently, \eqref{eq:kld_rl_i} can be written in vector form as:
\begin{align} \label{eq:kld_rl}
  C^{\eta, \theta} (p)
  &= (C_1^{\eta, \theta} (p) ~ \cdots ~ C_n^{\eta, \theta} (p))' \nonumber \\
  &= \textstyle \argmax_{z \in \mathbb X} ( z' p - \eta \mathcal D (z \,\|\, \theta) ),
\end{align}
where $\mathcal D (\cdot \,\|\, \cdot)$ denotes the \textit{Kullback-Leibler divergence}. As indicated by \eqref{eq:kld_rl}, the parameter $\eta$ controls a fundamental trade-off: when $\eta$ is large, $C^{\eta, \theta} (p)$ remains close to the reference distribution $\theta$; in contrast, when $\eta$ is small, strategy selection becomes more sensitive to the payoff vector $p$. Notably, both the Smith protocol \eqref{eq:smith_protocol} and the KLD-RL protocol \eqref{eq:kld_rl_i} do not depend on the population state $X^N(t)$.

Since each agent's revision follows an i.i.d. Poisson process, at most one agent can update its strategy at any given time. Consequently, the transition probability $P_{xy}^N (t)$ for $X^N(t)$ from $x \in \mathbb X^N$ to $y \in \mathbb X^N$ at a jump time $t$ of the Poisson processes is given by:
\begin{align} \label{eq:transition_probability}
  {\small
  P_{xy}^N(t) \!=\!
  \begin{cases}
    x_i \rho_{ij}(t) & \text{if } y \!=\! x \!+\! \frac{1}{N} (e_j \!-\! e_i), j \neq i \\
    1 \!-\! \sum_{i=1}^n \!\sum_{\substack{j=1 \\ j \neq i}}^n x_i \rho_{ij}(t) & \text{if } y \!=\! x \\
    0 & \text{otherwise},
  \end{cases}
  }
\end{align}
where $e_i$ is the $i$-th standard basis vector, and we use the shorthand notation $\rho_{ij}(t) = \rho_{ij}(p(t), X^N(t))$. As shown in \cite[Chapter~10]{Sandholm2010-SANPGA-2}, the stochastic process $X^N(t)$ can be approximated by the deterministic trajectory $x(t) \in \mathbb X$ governed by the following \textit{evolutionary dynamics model (EDM)}:
\begin{multline} \label{eq:edm}
  \textstyle \dot x_i(t) = \lambda \sum_{j=1}^n x_j(t) \rho_{ji} (p(t), x(t)) \\
  - \textstyle \lambda x_i(t) \sum_{j=1}^n \rho_{ij} (p(t), x(t)),
\end{multline}
where $\lambda$ is the rate parameter of the Poisson process, denoting the frequency at which each agent revises its strategy.

The accuracy of this approximation improves as the number of agents $N$ increases. In particular, as shown in \cite[Lemma~1]{6fda8e46-15d6-340f-bf79-5c3a826460f2}, when $N$ is sufficiently large and the payoff vector $p(t)$ is a function of $X^N(t)$, e.g., $p(t) = F(X^N(t))$, the following probabilistic bound holds:
\begin{multline} \label{eq:finite_population_approximation}
  \mathrm P \Big( \max_{t \in [0, T]} \|\widehat X^N(t) - x(t) \|_\infty < \epsilon \,\Big|\, X^N(0) = x_0 \Big) \\ \geq 1 - 2 (n-1) \exp (- \epsilon^2 c N),
\end{multline}
for some constant $c > 0$, and for any $\epsilon, T > 0$ and initial state $x_0 \in \mathbb X^N$. Here, $X^N(0) = x(0) = x_0$ and $\widehat X^N(t)$ denotes the piecewise linear interpolation of $X^N(t)$ (as detailed in Appendix~\ref{sec:interpolation}), and $x(t)$ is the solution to the deterministic EDM \eqref{eq:edm} with initial condition $x(0) = x_0$.

Inequality \eqref{eq:finite_population_approximation} provides a high-probability bound on the deviation between the interpolated process $\widehat X^N(t)$ and its deterministic counterpart $x(t)$ over the finite time interval $[0, T]$, with the approximation improving as $N$ increases. Given this result, a central research question in the literature is the stability analysis of the deterministic model \eqref{eq:edm}, which provides insight into the long-term behavior of the underlying finite-population process $X^N(t)$.

To model how $p(t)$ is determined in task allocation games \cite{10383344, 9561809}, we adopt the following dynamic model, where $p(t)$ evolves based on an internal state $q(t) = (q_1(t), \cdots, q_n(t))$ and the population state $X^N(t)$. For each task $i \in \{1, \cdots, m\}$, the model is given by:
\begin{subequations} \label{eq:task_allocation_games}
  \begin{align}
    \dot q_i(t) &= -\mathcal F_i(q_i(t), X^N(t)) + w_i \label{eq:task_allocation_games_a} \\
    p_i(t) &= G_i(q(t), X^N(t)). \label{eq:task_allocation_games_b}
  \end{align}
\end{subequations}
Here, $q_i(t) \in \mathbb R_{\geq 0}$ denotes the amount of remaining jobs in task~$i$. The function $\mathcal F_i: \mathbb R_{\geq 0} \times \mathbb X \to \mathbb R_{\geq 0}$, assumed to be continuously differentiable, represents the rate at which the agents complete jobs in task~$i$, depending on both the current job state $q_i(t)$ and the agents' strategy selection $X^N(t)$. To ensure nonnegativity of $q_i(t)$ for all $t \geq 0$, we impose $\mathcal F_i(0, x) = 0$ for all $x \in \mathbb X$ and $i \in \{1, \cdots, m\}$.
The positive constant $w_i$ represents the rate at which new jobs are added to task~$i$.

For simplicity, we consider the case where the number of tasks matches the number of available strategies ($m = n$) and assume that the payoff function in \eqref{eq:task_allocation_games_b} takes the form $G_i(q(t), X^N (t)) = q_i(t)$. This assumption simplifies the presentation while preserving the core intuition---agents are more likely to select strategies associated with tasks that have higher job demand.\footnote{For more general formulations, we refer the interested reader to \cite{10383344}.} We illustrate this model with the following example.
\begin{example} \label{example:task_allocation_games}
  As demonstrated in \cite{9561809}, the model described in  \eqref{eq:task_allocation_games} can be applied to a multi-robot resource collection scenario. In this setting, $n$ spatially distributed patches contain resources, with $q_i(t)$ denoting the volume of remaining resources at patch $i$. The function $\mathcal F_i$ denotes the rate at which agents collect resources from patch~$i$ and is given as:
  \begin{align} \label{eq:task_allocation_game_example}
    \mathcal F_i(q_i, X_i) = R_i \frac{e^{\alpha_i q_i} - 1}{e^{\alpha_i q_i} + 1} {X_i}^{\beta_i},
  \end{align}
  where $R_i, \alpha_i>0$ and $0 < \beta_i < 1$ are model parameters capturing saturation effects and diminishing returns.
\end{example}

\subsection{Passivity Techniques for Stability Analysis}
The concept of passivity \cite{schaft_springer} has been widely adopted in the game theory literature \cite{9029756, 9781277, 9219202, Fox2013Population-Game} to study the stability of agent decision-making processes in multi-agent games. Particularly relevant to our work are the notions of $\delta$-passivity and $\delta$-antipassivity, which apply to EDM \eqref{eq:edm} and the game model \eqref{eq:task_allocation_games}, respectively \cite{9029756}.\footnote{The symbol $\delta$ in $\delta$-passivity and $\delta$-antipassivity, as introduced in \cite{Fox2013Population-Game}, denotes \textit{differential} passivity in both the EDM and the task allocation game models.} Notably, EDM \eqref{eq:edm} under the KLD-RL revision protocol \eqref{eq:kld_rl_i} satisfies $\delta$-passivity with a positive surplus $\eta^\ast > 0$, as formally defined below. In contrast, the Smith protocol \eqref{eq:smith_protocol} satisfies $\delta$-passivity without surplus, i.e., $\eta^\ast = 0$.
\begin{definition} \label{def:delta_passivity_kld_rl}
  EDM \eqref{eq:edm} is said to be \textit{$\delta$-passive with surplus} $\eta^\ast > 0$ if there exists a continuously differentiable $\delta$-storage function $\mathcal S: \mathbb R^n \times \mathbb X \to \mathbb R_{\geq 0}$ such that, for all time-differentiable trajectories of $p(t)$ and for all $t \geq t_0 \geq 0$, the following inequality holds:
  \begin{align} \label{eq:delta_passivity_inequality}
    &\mathcal S(p(t), x(t)) - \mathcal S(p(t_0), x(t_0)) \leq \textstyle \int_{\tau = t_0}^t \big( \lambda^{-1} \dot p' (\tau) \dot x (\tau) \nonumber \\
    &\qquad\qquad - \lambda \eta^\ast \mathcal {V}'(p(\tau), x(\tau)) \mathcal V (p(\tau), x(\tau)) \big) \, \mathrm d\tau,
  \end{align}
  where $\mathcal V = (\mathcal V_1, \cdots, \mathcal V_n)$, and each component is defined by $\mathcal V_i (p, x) = \sum_{j=1}^n x_j \rho_{ji} (p, x) - x_i \sum_{j=1}^n \rho_{ij} (p, x)$.
\end{definition}
\vspace{.5em}

Note that EDM with the KLD-RL protocol satisfies inequality \eqref{eq:delta_passivity_inequality} with $\eta^\ast = \eta$.
In parallel, we define $\delta$-antipassivity for the game model \eqref{eq:task_allocation_games} as follows.
\begin{definition} \label{def:delta_antipassivity_task_allocation_game}
  The task allocation game \eqref{eq:task_allocation_games} is said to be \textit{$\delta$-antipassive} if there exists a continuously differentiable $\delta$-antistorage function $\mathcal L: \mathbb R_{\geq 0}^n \times \mathbb X \to \mathbb R_{\geq 0}$ such that, for all time-differentiable trajectories of $x(t)$ and for all $t \geq t_0 \geq 0$, the following inequality holds:
  \begin{align}
    \textstyle \mathcal L(q(t), x(t)) \!-\! \mathcal L(q(t_0), x(t_0)) \!\leq\! - \!\int_{\tau = t_0}^t \!\dot p' (\tau) \dot x (\tau) \, \mathrm d\tau.
  \end{align}
\end{definition}
\vspace{.5em}

As shown in \cite{10383344}, including the case of Example~\ref{example:task_allocation_games}, the function $\mathcal L$ can be explicitly constructed for a class of task allocation games. Throughout this work, we assume that the game model \eqref{eq:task_allocation_games} is $\delta$-antipassive. Furthermore, as discussed in \cite{9029756}, under the assumption of an infinite population of agents, i.e., $N$ is arbitrarily large, the feedback interconnection between a $\delta$-passive model \eqref{eq:edm} and a $\delta$-antipassive model \eqref{eq:task_allocation_games} guarantees the stability of the resulting closed-loop model. In particular, this interconnection guarantees convergence of $(q(t), x(t))$ to an equilibrium of the closed-loop model (see \cite[Theorems~4,5]{9029756} for a detailed discussion).

\begin{figure*} [t]
  \center
  \subfigure[]{
    \includegraphics[trim={0.15in 0.0in 0.11in 0.0in}, clip ,width=1.62in]{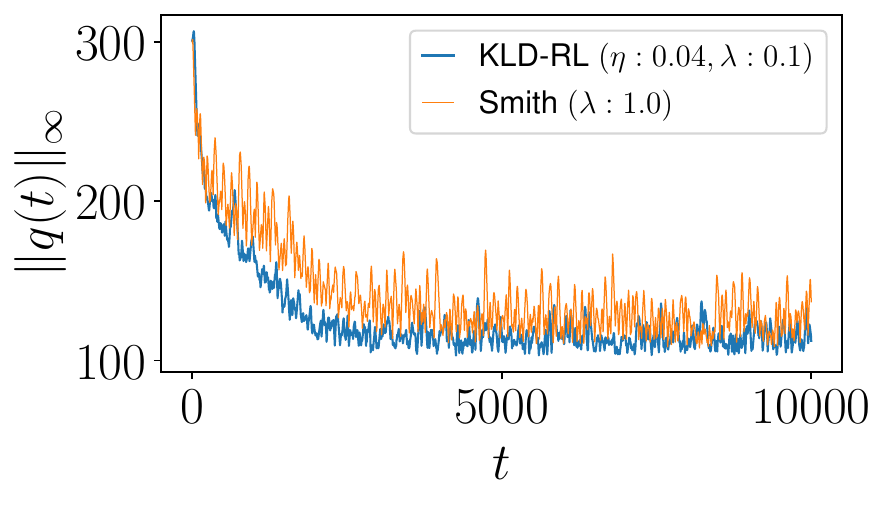}
    \label{fig:simulation_comparison_a}
  }
  \subfigure[]{
    \includegraphics[trim={0.15in 0.0in 0.12in 0.0in}, clip ,width=1.62in]{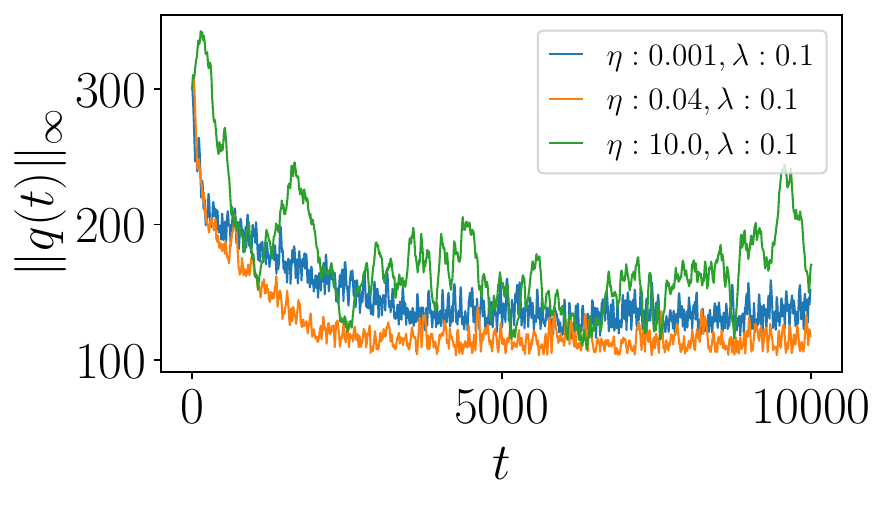}
    \label{fig:simulation_comparison_b}
  }
  \subfigure[]{
    \includegraphics[trim={0.15in 0.0in 0.11in 0.0in}, clip ,width=1.62in]{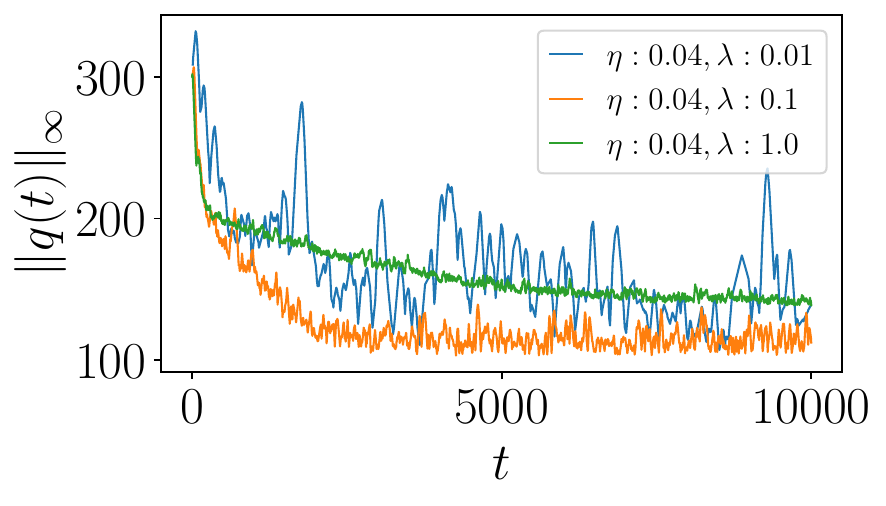}
    \label{fig:simulation_comparison_c}
  }
  \subfigure[]{
    \includegraphics[trim={0.15in 0.0in 0.12in 0.0in}, clip ,width=1.62in]{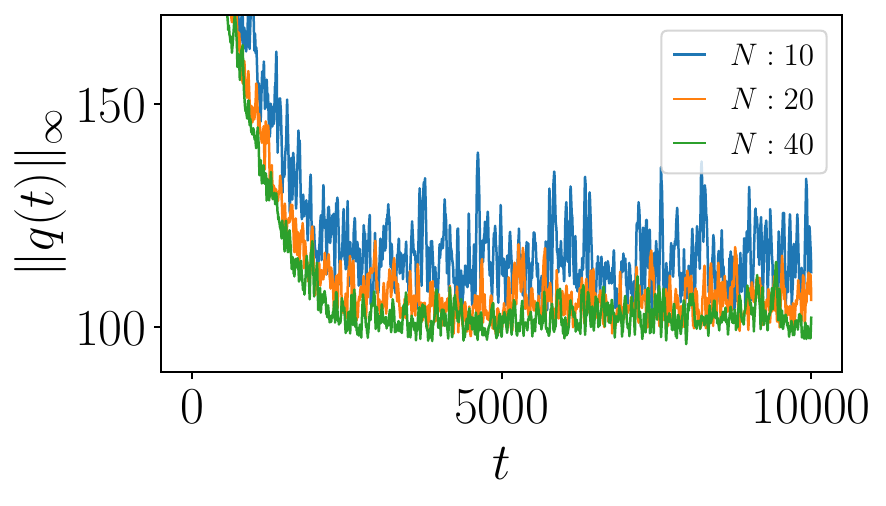}
    \label{fig:simulation_comparison_d}
  }
  \caption{(a) Comparison of $\| q(t) \|_\infty$ using the KLD-RL ($\eta=0.04, \lambda = 0.1$) and Smith protocols ($\lambda=1.0$) for $N=10$. (b) Trajectories of $\| q(t) \|_\infty$ under the KLD-RL protocol for $\eta=0.001, 0.04, 10.0$ with fixed $\lambda = 0.1$ and $N=10$. (c) Trajectories of $\| q(t) \|_\infty$ under the KLD-RL protocol for $\lambda=0.01, 0.1, 1.0$ with fixed $\eta = 0.04$ and $N=10$. (d) Trajectories of $\| q(t) \|_\infty$ under the KLD-RL protocol for $N = 10, 20, 40$ with fixed $\eta=0.04$ and $\lambda=0.1$.}
  \label{fig:simulation_comparison}
  \vspace{-1.0em}
\end{figure*}

\subsection{Problem Formulation}
In this work, we consider a setting where: $1)$ agents maintain an estimate $\hat p(t) \in \mathbb R^n$ of the payoff vector\footnote{For simplicity of presentation, we use $\hat p(t)$ to denote a common payoff vector estimate shared by all agents. However, as demonstrated in Section~\ref{sec:simulation} and formally established in \cite{10886571}, our analysis also extends to more general settings in which individual agents maintain distinct payoff vector estimates.}; $2)$ strategy revisions are subject to a constant time delay $d>0$; and $3)$ the population consists of a finite number of agents, implying that $X^N(t)$ and $x(t)$ are not identical, but may be related via the approximation in \eqref{eq:finite_population_approximation}. As a consequence, EDM \eqref{eq:edm} does not precisely describe the trajectory of $X^N(t)$, and the game state $q(t)$ in \eqref{eq:task_allocation_games_a} evolves based directly on $X^N(t)$.

To capture this problem setup, we introduce the following closed-loop model:
\begin{subequations} \label{eq:closed_loop_model_original}
  \begin{align}
    \dot q_i(t) &= -\mathcal F_i (q_i(t), X^N(t)) + w_i \\
    p_i(t) &= q_i(t) \\
    \dot {\widehat X}^N_i(t) &= \lambda (C_i^{\eta, \theta} (\hat p(t - d)) - {\widehat X}_i^N(t)) + \lambda \epsilon_i^N(t),
\end{align}
\end{subequations}
where $\widehat X_i^N(t)$ is the linear interpolation of $X_i^N(t)$ and $\epsilon_i^N (t)$ denotes an associated approximation error (as described in Appendix~\ref{sec:interpolation}). The term $C_i^{\eta, \theta} (\hat p(t - d))$ denotes the delayed strategy selection, incorporating a constant time delay~$d$ as formulated in \cite{10156113}.
We can equivalently express \eqref{eq:closed_loop_model_original} as follows:
\begin{subequations} \label{eq:closed_loop_model}
  \begin{align}
    \dot q_i(t) &= -\mathcal F_i (q_i(t), \widehat X^N (t)) + w_i + \tilde w_i^N (t) \label{eq:closed_loop_model_a} \\
    p_i(t) &= q_i(t) \label{eq:closed_loop_model_b} \\
    \dot {\widehat X}^N_i(t) &= \lambda (C_i^{\eta, \theta} (p(t)) \!-\! {\widehat X}_i^N(t)) \!+\! \lambda ( \epsilon_i^N(t) \!+\! \tilde v_i(t)), \label{eq:closed_loop_model_c}
\end{align}
\end{subequations}
The deviation between the true and interpolated states induces a modeling error given by:
\begin{align} \label{eq:noise_w}
  \tilde w_i^N(t) = \mathcal F_i(q_i(t), \widehat X^N (t)) - \mathcal F_i(q_i(t), X^N(t)).
\end{align}
Additionally, we define the noise induced by the estimation error as:
\begin{align} \label{eq:noise_v}
  \tilde v_i(t) = C_i^{\eta, \theta} (\hat p(t - d)) - C_i^{\eta, \theta} (p(t)).
\end{align}

In our analysis, we focus on the case where the game state $q(t)$ remains bounded and \eqref{eq:closed_loop_model} admits a unique equilibrium in the absence of noise. 
We formalize this as following assumption:
\begin{assumption} \label{assumption:bounded_game_state}
  The game state $q(t)$ in \eqref{eq:closed_loop_model} remains bounded over time. That is, there exists a positive constant $M_q$ such that $\|q(t)\|_\infty \leq M_q$ holds for all $t \geq 0$. Moreover, in the absence of noise, i.e., when $\epsilon_i^N(t) = \tilde w_i^N(t) = \tilde v_i(t) = 0$ for all $i$ in $\{1, \cdots, n\}$, the closed-loop model \eqref{eq:closed_loop_model} admits a unique equilibrium point $(q^\ast, x^\ast) \in \mathbb R_{\geq 0}^n \times \mathbb X$, where the reference distribution satisfies $\theta = x^\ast$.
\end{assumption}

The key research question we address is: \textit{How can the parameters $\lambda$ and $\eta$ be selected to minimize $\limsup_{t \to \infty} \| q(t) \|_\infty$---that is, to minimize the maximum remaining job demand across all tasks in the long run?} In Section~\ref{sec:analysis}, we analyze how these two parameters influence the behavior of the closed-loop model \eqref{eq:closed_loop_model}. Before delving into the analysis, we first present a simulation example to illustrate the main insights from our theoretical results.

\begin{remark}[On Assumption~\ref{assumption:bounded_game_state}]
We assume that the equilibrium point $x^\ast$ is known. In practice, this information may be unavailable without explicit knowledge of the game model \eqref{eq:task_allocation_games} or access to prior data. Unlike the Smith protocol, which serves as our baseline for performance comparison, the KLD-RL protocol \eqref{eq:kld_rl} requires a specified reference distribution $\theta$ in order to realize its full benefits. To assess the potential of the KLD-RL approach, we therefore assume in this paper that $x^\ast$ is known a priori.
\end{remark}

\section{Illustrative Example} \label{sec:simulation}
To illustrate the problem formulation and the influence of the parameters $\lambda$ and $\eta$ in \eqref{eq:closed_loop_model_c} on the trajectory of $\|q(t)\|_{\infty}$, we present a set of simulation results. Specifically, we consider the task allocation game defined by \eqref{eq:task_allocation_game_example} with $n=3$ strategies. The game model's parameters are set as $R_1=R_2=R_3 = 3.44$, $\alpha_1 = \alpha_2 = \alpha_3 = 0.036$, and $\beta_1 = \beta_2 = \beta_3 = 0.91$. The growth rate vector in \eqref{eq:closed_loop_model_original} is chosen as $w = (0.5, 1, 2)$, and the time delay $d$ in the strategy revision is set to $d = 10$. Under Assumption~\ref{assumption:bounded_game_state}, let $(q^\ast, x^\ast) \in \mathbb R_{\geq 0}^n \times \mathbb X$ denote the unique equilibrium point of \eqref{eq:closed_loop_model} in the absence of noise, where the reference distribution satisfies $\theta = x^\ast$. In this case, $x^\ast \approx (0.13, 0.28, 0.59)$ and the corresponding $\| q^\ast \|_\infty$ represents the minimum attainable value of $\limsup_{t \to \infty} \|q(t)\|_{\infty}$.

Each agent~$k$ maintains an estimate $\hat p^{(k)} (t)$ of the payoff vector, which is updated at discrete time steps $t = 0, 1, \cdots$. If agent~$k$ is designated as an \textit{observer}, it has direct access to the true payoff vector, i.e., $\hat p^{(k)} (t) = p(t)$ for all $t = 0, 1, \cdots$. Otherwise, it updates its estimate using a consensus-based rule at each time step $t = 0, 1, \cdots$, according to:
\begin{align}
  \hat p^{(k)} (t) = \textstyle\frac{1}{|\mathbb N_k|} \sum_{l \in \mathbb N_k} \hat p^{(l)} (t),
\end{align}
where $\mathbb N_k$ denotes the set of neighbors of agent~$k$ in the underlying communication graph. The communication graph is constructed as a strongly connected Erd{\H o}s-R{\' e}nyi random graph with a connection probability of $0.2$. One-tenth of the agents are randomly selected as observers; for example, in a simulation with $N=10$ agents, a single agent is designated as an observer with access to $p(t)$. All agents initialize their payoff vector estimates as $\hat p^{(k)}(0) = (0, 0, 0)$.

Simulations are conducted with a population size of $N = 10, 20, 40$. The initial condition is set to $q(0) = (100, 200, 300)$, and each agent's initial strategy is selected uniformly at random from the three available strategies. Figure~\ref{fig:simulation_comparison} presents the simulation results under various parameter settings. As shown in Fig.~\ref{fig:simulation_comparison_a}, where the parameter $\lambda$ for the Smith protocol is chosen to minimize $\limsup_{t \to \infty} \| q(t) \|_\infty$ over a range of candidate values, the KLD-RL protocol \eqref{eq:kld_rl} with $\theta = x^\ast$ demonstrates greater robustness to noise compared to the Smith protocol \eqref{eq:smith_protocol} with $\varrho = 1/(n-1) M_q$, indicating that the KLD-RL protocol is better suited to our problem formulation. Figure~\ref{fig:simulation_comparison_b} illustrates that increasing $\eta$ while keeping $\lambda$ fixed in the KLD-RL protocol reduces $\limsup_{t \to \infty} \| q(t) \|_\infty$. However, overly large values of $\eta$ tend to induce increased variance in the trajectory of $\| q(t) \|_\infty$. Similarly, Fig.~\ref{fig:simulation_comparison_c} shows that decreasing $\lambda$ in the KLD-RL protocol yields a similar trade-off. Additionally, Fig.~\ref{fig:simulation_comparison_d} demonstrates that the overall performance of the closed-loop---in terms of achieving lower values of $\limsup_{t \to \infty} \|q(t)\|_\infty$ and reduced variance---improves as the population size $N$ increases.

\section{Analysis of the KLD-RL Protocol} \label{sec:analysis}
We analyze how the parameters $\lambda$ and $\eta$ influence the long-term behavior of the game state $q(t)$. We begin by summarizing the key implications of our analytical results:
\begin{itemize}
\item \textit{Effect of $\lambda$}: When $\epsilon_i^N(t)$ and $\tilde v_i(t)$ are fixed, the magnitude of the noise term $\lambda(\epsilon_i^N(t) + \tilde v_i(t))$ in the closed-loop model~\eqref{eq:closed_loop_model_c} scales linearly with $\lambda$. Hence, decreasing $\lambda$ would reduce the noise level and promote convergence to equilibrium. However, as shown in Theorem~\ref{theorem:effect_of_noise}, a smaller $\lambda$ may also amplify the overall impact of noise on agents' strategy selection. This effect can be alleviated through proper tuning of $\eta$, which helps counterbalance the increased sensitivity introduced by a reduced $\lambda$.

\item \textit{Effect of $\eta$}: When the population size $N$ is finite, setting $\eta$ too large may excessively constrain the responsiveness of agents' strategy selection process to changes in the game state $q(t)$, as supported by Proposition~\ref{proposition:mean_variance_of_stationary_distribution}.
\end{itemize}
These analytical insights are supported by the simulation results presented in Figs.~\ref{fig:simulation_comparison_b} and \ref{fig:simulation_comparison_c}.

To study how the parameters $\lambda$ and $\eta$ affect the long-term behavior of $q(t)$, we consider the closed-loop model \eqref{eq:closed_loop_model} under Assumption~\ref{assumption:bounded_game_state}. Let $(q^\ast, x^\ast)$ denote the unique equilibrium point of \eqref{eq:closed_loop_model} in the absence of noise, where the reference distribution satisfies $\theta = x^\ast$. At this equilibrium, $\| q^\ast \|_\infty$ corresponds to the minimum achievable value of $\limsup_{t \to \infty} \|q(t)\|_{\infty}$, and $x^\ast$ corresponds to the optimal stationary distribution over the available strategies. Given that the game model is $\delta$-antipassive, \cite[Theorems~5]{9029756} ensures that the state of the closed-loop \eqref{eq:closed_loop_model}, with $\theta=x^\ast$, converges to $(q^\ast, x^\ast)$ in the absence of noise.

Throughout our analysis, we assume that the KLD-RL protocol \eqref{eq:kld_rl_i} in the closed-loop model \eqref{eq:closed_loop_model} satisfies $\theta = x^\ast$. Under this setting, the equality $C^{\eta, \theta} (p(t)) - X^N(t) = 0$ implies that agents are selecting the strategies that maximize the regularized average payoff $p'(t) X^N(t) - \eta \mathcal D (X^N(t) \,\|\, x^\ast)$. This reflects a preference for tasks with the highest remaining jobs---as captured by $p(t) = q(t)$---while incorporating a regularization term that penalizes deviation from the optimal strategy distribution $x^\ast$. In addition, when the game model satisfies $\mathcal F(q(t), X^N(t)) = w$, the state $X^N(t)$ matches the equilibrium $x^\ast$ for any $\eta > 0$, and $\|q(t)\|_\infty$ attains its minimum value at $\| q^\ast \|_\infty$. Therefore, the deviation of $X^N(t)$ from $C^{\eta, \theta} (p(t))$ can be interpreted, in part, as a key indicator of how effectively agents are adapting the strategies toward the optimal strategy distribution $x^\ast$.

The following theorem characterizes how the parameters $\lambda$ and $\eta$ influence the effect of the noise terms $\tilde w^N(t)$, $\epsilon^N(t)$, and $\tilde v(t)$ on the convergence of the interpolated population state $\widehat X^N(t)$ to the best response $C^{\eta, \theta} (p(t))$.

\begin{theorem} \label{theorem:effect_of_noise}
  Consider the closed-loop model \eqref{eq:closed_loop_model}, and recall the $\delta$-storage function $\mathcal S$ and $\delta$-antistorage function $\mathcal L$, defined for EDM \eqref{eq:edm} and the game model \eqref{eq:task_allocation_games}, respectively. For any time horizon $T>0$, the following inequality holds:
  \begin{align} \label{eq:effect_of_noise}
    &\textstyle \int_0^T \| C^{\eta,\theta}(p(t)) - \widehat X^N (t) \|_2^2 \, \mathrm dt \nonumber \\ 
    &\textstyle \leq \! \frac{1}{\lambda^2 \eta} \! \left( \alpha_\lambda \!+\! \int_0^T \! | g_\lambda (p(t), \widehat X^N (t), \tilde w^N (t), \epsilon^N(t), \tilde v(t)) | \mathrm dt \right),
  \end{align} 
  where $\alpha_\lambda = \lambda \mathcal S(p(0), \widehat X^N (0)) + \mathcal L (q(0), \widehat X^N (0))$ and the function $g_\lambda$ is defined as $g_\lambda (p, \widehat X^N, \tilde w^N, \epsilon^N, \tilde v) = - 2 \lambda (\epsilon^N + \tilde v)' \tilde w^N + \lambda (\lambda \nabla_x \mathcal S (p, \widehat X^N) + \mathcal F(q, \widehat X^N) - w )' (\epsilon^N + \tilde v)  + ( \nabla_q \mathcal L(q, \widehat X^N) - \lambda \mathcal V^{\eta, \theta} (p, \widehat X^N) )' \tilde w^N$ with $\mathcal V^{\eta, \theta} (p, \widehat X^N) = C^{\eta, \theta}(p) - \widehat X^N$. Additionally, the function $C^{\eta, \theta}$ is Lipschitz continuous, which leads to the following bound:
  \begin{align} \label{eq:choice_function_lipschitz_continuity}
    \|\tilde v(t) \|_2 \leq \eta^{-1} \|p(t) - \hat p(t-d)\|_2,
  \end{align}
  implying that $\|\tilde v(t) \|_2 \to 0$ as $\eta \to \infty$, provided the payoff vector estimation error $\|p(t) - \hat p(t-d)\|_2$ remains bounded.
\end{theorem}

The proof of the theorem is presented in Appendix~B. The inequality in \eqref{eq:effect_of_noise} shows how noise, captured by the function $g_\lambda$, affects the deviation between $\widehat X^N(t)$ and $C^{\eta, \theta} (p(t))$. The term $(\lambda^2 \eta)^{-1}$ acts as an upper bound on this influence, meaning that decreasing $\lambda$ or $\eta$ can amplify the closed-loop model's sensitivity to noise. The function $g_\lambda$ can be upper-bounded by:
\begin{multline*}
  g_\lambda (p, X^N, \tilde w^N, \epsilon^N, \tilde v) \leq 2 \lambda \| \epsilon^N+\tilde v\|_2 \|\tilde w^N \|_2 \\ + \lambda (\lambda M_{\nabla_x \mathcal S} + M_{\mathcal F} + \|w\|_2 ) \|\epsilon^N + \tilde v \|_2 + (  M_{\nabla_q \mathcal L} + \lambda M_{\mathcal V} ) \|\tilde w^N \|_2,
\end{multline*}
where $M_{\nabla_x \mathcal S}$, $M_{\mathcal F}$, $M_{\nabla_q \mathcal L}$, and $M_{\mathcal V}$ are positive constants that bound $\nabla_x \mathcal S$, $\mathcal F$, $\nabla_q \mathcal L$, and $\mathcal V^{\eta, \theta}$, respectively.\footnote{Given Assumption~\ref{assumption:bounded_game_state} and the boundedness of $\mathbb X$, such constants are guaranteed to exist.} Notably, although reducing $\lambda$ lowers this upper bound, it also increases the term $(\lambda^2 \eta)^{-1}$, thereby amplifying the effect of noise in the closed-loop model. To balance this trade-off, increasing $\eta$ can help compensate for the smaller $\lambda$. In particular, a larger $\eta$ not only reduces $(\lambda^2 \eta)^{-1}$, but also directly decreases $\tilde v(t)$, due to the Lipschitz continuity of $C^{\eta, \theta}$, as shown in \eqref{eq:choice_function_lipschitz_continuity}, provided that $\hat p(t-d)$ remains close to $p(t)$.

The remaining question is: \textit{Can agents increase $\eta$ arbitrarily to compensate for a smaller $\lambda$, thereby further mitigating the influence of noise on strategy selection?}


Recall that $x^\ast$ denotes the unique equilibrium of \eqref{eq:closed_loop_model} in the absence of noise. 
Under Assumption~\ref{assumption:bounded_game_state}, the function $C^{\eta, \theta} (p(t))$ converges to $x^\ast$ as $\eta$ becomes sufficiently large. In the analysis that follows, we show that when $\eta$ is excessively large; hence, we approximate $C^{\eta, \theta} (p(t))$ with $x^\ast$, the evolution of $X^N(t)$ becomes independent of the payoff vector $p(t)$, effectively rendering \eqref{eq:closed_loop_model} open-loop. As a result, \eqref{eq:closed_loop_model_c} becomes decoupled from \eqref{eq:closed_loop_model_a}. To facilitate the analysis, we consider the case with no time delay in strategy revision (i.e., $d=0$).

With $C^{\eta, \theta} (p(t)) = x^\ast$, the transition probability \eqref{eq:transition_probability} simplifies to:
\begin{align} \label{eq:transition_probability_large_eta}
  P_{xy}^N 
  &= \begin{cases}
    x_i x_j^\ast & \text{if } y = x + \frac{1}{N} (e_j - e_i), ~ j \neq i \\
    \sum_{i=1}^n x_i x_i^\ast & \text{if } y = x \\
    0 & \text{otherwise}.
  \end{cases}
\end{align}
The long-term behavior of $X^N(t)$ is governed by the stationary distribution $\mu^N = (\mu_x^N)_{x \in \mathbb X^N}$, which satisfies the standard balance equation: $\mu_x^N = \sum_{y \in \mathbb X^N} \mu_y^N P_{yx}^N$ for $x \in \mathbb X^N$. The following proposition characterizes the mean and variance of $X^N(t)$ under the
stationary distribution.
\begin{proposition} \label{proposition:mean_variance_of_stationary_distribution}
  Consider the population state $X^N(t)$ whose transition probability is given as in \eqref{eq:transition_probability_large_eta}. Then, under the stationary distribution, the mean $\mathrm E(X^N(t))$ and the variance $\sum_{i=1}^n \mathrm{Var}(X_i^N(t))$ are, respectively, given by:
  \begin{subequations}
    \begin{align}
      \mathrm E (X^N (t)) &= x^\ast \\
      \textstyle\sum_{i=1}^n \mathrm{Var}(X_i^N(t)) &= N^{-1} (1 - {x^\ast}'x^\ast).                               
    \end{align}
  \end{subequations}
\end{proposition}
\vspace{.5em}

The proof of the proposition is provided in Appendix~C. Proposition~\ref{proposition:mean_variance_of_stationary_distribution} indicates that when $\eta$ is excessively large, the population state $X^N(t)$ becomes independent of $p(t)$ and evolves according to $\mu^N$, whose variance increases as the population size $N$ decreases. As shown in Fig.~\ref{fig:simulation_comparison_b}, in this regime the agents cannot adaptively switch strategies based on $p(t)$, leading to overshooting in the early stages. Furthermore, the variance of $X^N(t)$ may substantially affect the variability of $\|q(t)\|$ in later periods. Taken together with Theorem~\ref{theorem:effect_of_noise}, the proposition highlights the importance of carefully tuning $\lambda$ and $\eta$ to improve both noise robustness and convergence performance.

\section{Conclusions} \label{sec:conclusions}
We investigated the robustness of the KLD-RL model in finite-population games, where agents choose strategies based on noisy payoff estimates and update their strategies with a time delay. We analyzed how two key parameters affect the sensitivity of strategy selection to noise. Our results highlight how tuning these parameters enables agents to effectively learn the optimal equilibrium in task allocation games.
For future work, we aim to develop data-driven methods for optimizing the model’s parameters. In this study, we assumed that the optimal population state $x^\ast$ is known and set the model’s reference distribution to $\theta = x^\ast$. To relax this assumption, we plan to explore approaches for learning $x^\ast$ directly and identifying an appropriate choice for $\theta$.

\appendix

\subsection{Linear Interpolation $\widehat X^N(t)$ of Population State $X^N(t)$} \label{sec:interpolation}
Let $\mathbb T$ denote the set of arrival times of the Poisson processes that govern the timing of strategy revisions for all agents. For any $t \in [t_1, t_2)$, where $t_1, t_2 \in \mathbb T$ are two consecutive arrival times with $t_1 < t_2$, we define a piecewise linear interpolation $\widehat X^N (t)$ of $X^N(t)$ as follows:
\begin{align} \label{eq:x_hat_definition}
  \widehat X^N(t) = X^N(t_1) + \frac{t - t_1}{t_2-t_1} \left( X^N (t_2) - X^N (t_1) \right).
\end{align}
Hence, for each $i \in \{1, \cdots, n\}$ and $t \in (t_1, t_2)$, we can deduce that\footnote{By definition \eqref{eq:x_hat_definition}, $\widehat X^N(t)$ is differentiable at all $t \notin \mathbb T$.}
\begin{align}
  \dot {\widehat{X}}_i^N(t)
  &= \frac{1}{t_2-t_1} \left( X_i^N (t_2) - X_i^N (t_1) \right) \nonumber \\
  &= \lambda (C_i^{\eta, \theta} (\hat p(t-d)) - {\widehat X}_i^N(t)) + \lambda \epsilon_i^N(t),
\end{align}
where $\epsilon_i^N(t) = \frac{1}{\lambda(t_2-t_1)} (X_i^N (t_2) - X_i^N (t_1)) - (C_i^{\eta, \theta} (\hat p(t-d)) - {\widehat X}_i^N(t))$, which is bounded for all $t \in (t_1, t_2)$.

 \subsection{Proof of Theorem~\ref{theorem:effect_of_noise}}
 We provide a two-part proof.

 \paragraph{Derivation of \eqref{eq:effect_of_noise}}
 The $\delta$-passivity with surplus $\eta$ (see Definition~\ref{def:delta_passivity_kld_rl}) of EDM \eqref{eq:edm} implies that the $\delta$-storage function $\mathcal S$ satisfies the following relations for all $(p,x) \in \mathbb R^n \times \mathbb X$:
 \begin{subequations} \label{eq:delta_passivity_algebraic}
   \begin{align}
     &\nabla_x' \mathcal S (p, x) \mathcal V^{\eta, \theta} (p, x) \leq - \eta \mathcal {V^{\eta, \theta}}' (p, x) \mathcal V^{\eta, \theta} (p, x) \\
     &\nabla_p \mathcal S (p, x) = \mathcal V^{\eta, \theta} (p, x),
   \end{align}
 \end{subequations}
 where $\mathcal V^{\eta, \theta} (p,x) = C^{\eta, \theta} (p) - x$.
 Additionally, by the $\delta$-antipassivity of the game model \eqref{eq:task_allocation_games} (see Definition~\ref{def:delta_antipassivity_task_allocation_game}), the $\delta$-antistorage function $\mathcal L$ satisfies, for all $(q, x) \in \mathbb R_{\geq 0}^n \times \mathbb X$:
 \begin{subequations} \label{eq:delta_antipassivity_algebraic}
   \begin{align} 
     &\nabla_q' \mathcal L (q, x) (-\mathcal F(q, x) + w) \leq 0 \label{eq:delta_antipassivity_algebraic_a} \\
     &\nabla_x \mathcal L(q,x) = \mathcal F(q,x) - w.
   \end{align}
 \end{subequations}

 Consider time-varying trajectories $q(t) \in \mathbb R_{\geq 0}^n$ and $x(t) \in \mathbb X$, which evolve according to the closed-loop model \eqref{eq:closed_loop_model}, with $x(t) = \widehat X^N(t)$. Using \eqref{eq:delta_passivity_algebraic} and \eqref{eq:delta_antipassivity_algebraic}, the time derivative of the composite Lyapunov-like function $\lambda \mathcal S(p, x) + \mathcal L (q, x)$ is computed as:\footnote{For brevity, we omit the time variable $t$ in the derivation.}
 \begin{align*}
   &\frac{\mathrm d}{\mathrm dt} \left( \lambda \mathcal S(p, x) + \mathcal L (q, x) \right) \\
   &= \lambda \nabla_x' \mathcal S(p, x) (\lambda \mathcal V^{\eta, \theta} (p, x) + \lambda (\epsilon^N+\tilde v) ) + \lambda \nabla_p' \mathcal S(p, x) \dot p \\
   &\qquad + \nabla_q' \mathcal L(q, x) (-\mathcal F(q, x) + w + \tilde w^N) + \nabla_x' \mathcal L(q,x) \dot x \\
   &= \lambda^2 \nabla_x' \mathcal S(p, x) \mathcal V^{\eta, \theta} (p, x) + \lambda^2 \nabla_x' \mathcal S(p, x) (\epsilon^N+\tilde v)  \\
   &\qquad + (\dot x - \lambda (\epsilon^N+\tilde v))' \dot p + \nabla_q' \mathcal L(q, x) (-\mathcal F(q, x) + w) \nonumber \\
   &\qquad + \nabla_q' \mathcal L(q, x) \tilde w^N - (\dot p + \tilde w^N)' \dot x \\
   &= \lambda^2 \underbrace{\nabla_x' \mathcal S(p, x) \mathcal V^{\eta, \theta} (p, x)}_{\leq -\eta \mathcal {V^{\eta, \theta}}' (p,x) \mathcal V^{\eta, \theta} (p,x)} + \underbrace{\nabla_q' \mathcal L(q, x) (-\mathcal F(q, x) + w)}_{\leq 0} \\
   &\qquad + \lambda (\lambda \nabla_x \mathcal S(p, x) + \mathcal F(q,x) - w )' (\epsilon^N+\tilde v) \\
   &\qquad + (\nabla_q \mathcal L(q, x) - \lambda \mathcal V^{\eta, \theta}(p,x))' \tilde w^N - 2 \lambda (\epsilon^N+\tilde v)' \tilde w^N.
 \end{align*}

 Let us define $\alpha_\lambda = \lambda \mathcal S(p(0), x(0)) + \mathcal L (q(0), x(0))$. Then, we obtain the following inequality:
 \begin{align*}
   \frac{\int_0^T \mathcal {V^{\eta, \theta}}' (p(t),x(t)) \mathcal V^{\eta, \theta} (p(t),x(t)) \,\mathrm dt}{\alpha_\lambda+\int_0^T |g_\lambda (p(t), x(t), \tilde w^N(t), \epsilon^N(t), \tilde v(t))| \,\mathrm dt}  \leq (\lambda^2 \eta)^{-1},
 \end{align*}
 where $g_{\lambda}$ is defined as in the statement of the theorem. Substituting $x(t)$ with $\widehat X^N(t)$ completes the derivation.

 \paragraph{Derivation of \eqref{eq:choice_function_lipschitz_continuity}}
 Following the analysis in \cite{gao2018propertiessoftmaxfunctionapplication}, define the log-sum-exp function $\mathrm{lse}^{\eta, \theta}: \mathbb R^n \to \mathbb R$ associated with \eqref{eq:kld_rl_i} as:
 \begin{align*}
   \mathrm{lse}^{\eta, \theta} (p) = \textstyle \eta \ln \left( \sum_{j=1}^n \theta_j \exp (\eta^{-1} p_j) \right), ~ p \in \mathbb R^n.
 \end{align*}
 The first- and second-order partial derivatives of $\mathrm{lse}^{\eta, \theta} (p)$ with respect to $p$ are given by:
 \begin{align*}
   \frac{\partial}{\partial p_i} \mathrm{lse}^{\eta, \theta} (p) = \frac{\theta_i \exp (\eta^{-1} p_i)}{\sum_{j=1}^n \theta_j \exp (\eta^{-1} p_j)} = C_i^{\eta, \theta}(p)
 \end{align*}
 and
 \begin{multline*}
   \frac{\partial^2}{\partial p_i^2} \mathrm{lse}^{\eta, \theta} (p) = \frac{\eta^{-1} \theta_i \exp (\eta^{-1} p_i) }{\sum_{j=1}^n \theta_j \exp (\eta^{-1} p_j)} \\
   - \frac{\eta^{-1} \theta_i^2 \exp (2 \eta^{-1} p_i)}{\left( \sum_{j=1}^n \theta_j \exp (\eta^{-1} p_j) \right)^2}
 \end{multline*}
 \begin{align*}
   \frac{\partial^2}{\partial p_k \partial p_i} \mathrm{lse}^{\eta, \theta} (p) \!=\! \frac{-\eta^{-1} \theta_k \theta_i \exp (\eta^{-1} p_k) \exp (\eta^{-1} p_i)}{\left( \sum_{j=1}^n \theta_j \exp (\eta^{-1} p_j) \right)^2}, \forall i \neq k.
 \end{align*}
 Hence, the gradient and Hessian of $\mathrm{lse}^{\eta, \theta} (p)$, respectively, become
 \begin{align*}
   \nabla \mathrm{lse}^{\eta, \theta} (p) = C^{\eta, \theta}(p)
 \end{align*}
 and
 \begin{align*}
   \nabla^2 \mathrm{lse}^{\eta, \theta} (p) = \eta^{-1} \left( \mathrm{diag} (C^{\eta, \theta} (p)) - C^{\eta, \theta} (p) {C^{\eta, \theta}}'(p) \right),
 \end{align*}
 where $\mathrm{diag} (C^{\eta, \theta} (p))$ denotes the diagonal matrix whose diagonal entries are given by the components of $C^{\eta, \theta} (p)$. Consequently, by following similar steps as in the proof of \cite[Proposition~4]{gao2018propertiessoftmaxfunctionapplication}, it holds for any $p,v \in \mathbb R^n$ that
 \begin{align*}
   0 \leq v' \nabla^2 \mathrm{lse}^{\eta, \theta} (p) v \leq \eta^{-1} \|v\|_2^2.
 \end{align*}
 Using \cite[Lemma~5]{gao2018propertiessoftmaxfunctionapplication}, we conclude the function $C^{\eta, \theta}$ is Lipschitz continuous:
 \begin{align*}
   \| C^{\eta, \theta} (p) - C^{\eta, \theta} (\hat p) \|_2 \leq \eta^{-1} \|p - \hat p\|_2,
 \end{align*}
 which yields \eqref{eq:choice_function_lipschitz_continuity}.
 \hfill\QED

 \subsection{Proof of Proposition~\ref{proposition:mean_variance_of_stationary_distribution}}
 To compute the expectation $\mathrm E(X^N(t))$, using $\mu_y^N = \sum_{x \in \mathbb X^N} \mu_x^N P_{xy}^N$, we proceed as follows:
 \begin{align*}
   &\sum_{y \in \mathbb X^N} y \mu_y^N \nonumber \\
   &= \sum_{y \in \mathbb X^N} \sum_{x \in \mathbb X^N} y \mu_x^N P_{xy}^N \nonumber \\
   &= \sum_{x \in \mathbb X^N} \! x \mu_x^N \sum_{i=1}^n x_i x_i^\ast \!+\! \sum_{x \in \mathbb X^N} \sum_{j=1}^n \sum_{\substack{i=1 \\ i \neq j}}^n (x \!+\! \frac{1}{N} (e_j \!-\! e_i)) \mu_{x}^N x_i x_j^\ast \nonumber \\
   &= \sum_{x \in \mathbb X^N} \! x \mu_x^N \underbrace{\sum_{i=1}^n \sum_{j=1}^n x_i x_j^\ast}_{=1} \!+\! \frac{1}{N} \sum_{x \in \mathbb X^N} \sum_{j=1}^n \sum_{\substack{i=1 \\ i \neq j}}^n (e_j \!-\! e_i) \mu_{x}^N x_i x_j^\ast
 \end{align*}
 which implies
 \begin{align*}
   \sum_{x \in \mathbb X^N} \sum_{i=1}^n \sum_{j=1}^n e_j \mu_{x}^N x_i x_j^\ast &= \sum_{x \in \mathbb X^N} \sum_{i=1}^n \sum_{j=1}^n e_i \mu_{x}^N x_i x_j^\ast.
 \end{align*}
 Consequently, we conclude that
 \begin{align}
   \sum_{x \in \mathbb X^N} \mu_{x}^N x^\ast = \sum_{x \in \mathbb X^N} \mu_{x}^N x \implies x^\ast = \mathrm E (X^N(t)).
 \end{align}

 Next, to compute $\sum_{i=1}^n \mathrm{Var}(X_i^N(t))$, we analyze:
 \begin{align*}
   &\sum_{y \in \mathbb X^N} (y - x^\ast)'(y - x^\ast) \mu_y^N \nonumber \\
   &= \sum_{y \in \mathbb X^N} \sum_{x \in \mathbb X^N} (y - x^\ast)'(y - x^\ast) \mu_x^N P_{xy}^N \nonumber \\
   &= \sum_{x \in \mathbb X^N} (x - x^\ast)'(x - x^\ast) \mu_x^N \sum_{i=1}^n x_i x_i^\ast \nonumber \\
   &\qquad + \sum_{x \in \mathbb X^N} \sum_{j=1}^n \sum_{\substack{i=1 \\ i \neq j}}^n (x - x^\ast + \frac{1}{N} (e_j - e_i))' \nonumber \\
   &\qquad\qquad\qquad\qquad\qquad (x - x^\ast + \frac{1}{N} (e_j - e_i)) \mu_{x}^N x_i x_j^\ast \nonumber \\
   &= \sum_{x \in \mathbb X^N} (x - x^\ast)'(x - x^\ast) \mu_x^N \nonumber \\
   &\qquad + \frac{1}{N^2} \sum_{x \in \mathbb X^N} \sum_{i=1}^n \sum_{\substack{j=1 \\ j \neq i}}^n \underbrace{(e_j - e_i)'(e_j - e_i)}_{=2} \mu_{x}^N x_i x_j^\ast \nonumber \\
   &\qquad + \frac{2}{N} \sum_{x \in \mathbb X^N} \sum_{i=1}^n \sum_{j=1}^n (x - x^\ast)' (e_j - e_i) \mu_{x}^N x_i x_j^\ast \nonumber \\
   &= \sum_{x \in \mathbb X^N} (x - x^\ast)'(x - x^\ast) \mu_x^N + \frac{2}{N^2} \sum_{x \in \mathbb X^N} (1 - \sum_{i=1}^n x_i x_i^\ast) \mu_{x}^N \nonumber \\
   &\qquad + \frac{2}{N} \sum_{x \in \mathbb X^N} \sum_{i=1}^n \sum_{j=1}^n (x - x^\ast)' (e_j - e_i) \mu_{x}^N x_i x_j^\ast
 \end{align*}
 which implies
 \begin{multline*}
   \frac{2}{N} \sum_{x \in \mathbb X^N} (x - x^\ast)' x \mu_{x}^N - \frac{2}{N} \sum_{x \in \mathbb X^N}  (x - x^\ast)' x^\ast \mu_{x}^N \\
   = \frac{2}{N^2} (1 - {x^\ast}'x^\ast)
 \end{multline*}
 and we conclude that
 \begin{align*}
   &\sum_{x \in \mathbb X^N} (x - x^\ast)'(x - x^\ast) \mu_{x}^N = \frac{1}{N} (1 - {x^\ast}'x^\ast) \\
   \implies& \sum_{i=1}^n \mathrm{Var}(X_i^N(t)) = \frac{1}{N} (1 - {x^\ast}'x^\ast).
 \end{align*}
 This completes the proof. \hfill\QED

 \balance
\bibliographystyle{IEEEtran}
\bibliography{IEEEabrv,references}

\end{document}